
\def\figflag{y}
\input epsf
\def\boringfonts{y}  

%
%
\def\unredoffs{} \def\redoffs{\voffset=-.31truein\hoffset=-13truemm}
\def\speclscape{}
%
%
%
%
%
\newbox\leftpage \newdimen\fullhsize \newdimen\hstitle \newdimen\hsbody
\tolerance=1000\hfuzz=2pt
\catcode`\@=11 
\def\bigans{b }
\def\answ{b } 
\ifx\answ\bigans\message{(This will come out unreduced.}
\magnification=1200\unredoffs\baselineskip=16pt plus 2pt minus 1pt
\hsbody=\hsize \hstitle=\hsize 
\else\message{(This will be reduced.} \let\l@r=L
\magnification=1000\baselineskip=16pt plus 2pt minus 1pt \vsize=7truein
\redoffs \hstitle=8truein\hsbody=4.75truein\fullhsize=10truein\hsize=\hsbody
\output={\ifnum\pageno=0 
  \shipout\vbox{\speclscape{\hsize\fullhsize\makeheadline}
    \hbox to \fullhsize{\hfill\pagebody\hfill}}\advancepageno
  \else
  \almostshipout{\leftline{\vbox{\pagebody\makefootline}}}\advancepageno 
  \fi}
\def\almostshipout#1{\if L\l@r \count1=1 \message{[\the\count0.\the\count1]}
      \global\setbox\leftpage=#1 \global\let\l@r=R
 \else \count1=2
  \shipout\vbox{\speclscape{\hsize\fullhsize\makeheadline}
      \hbox to\fullhsize{\box\leftpage\hfil#1}}  \global\let\l@r=L\fi}
\fi
%
\newcount\yearltd\yearltd=\year\advance\yearltd by -1900

\def\Title#1#2{\nopagenumbers\abstractfont\hsize=\hstitle\rightline{#1}%
\vskip 1in\centerline{\titlefont #2}\abstractfont\vskip .5in\pageno=0}
\def\Date#1{\vfill\leftline{#1}\tenpoint\supereject\global\hsize=\hsbody%
\footline={\hss\tenrm\folio\hss}}
%

\def\draftmode{\message{ DRAFTMODE }\def\draftdate{{\rm preliminary draft:
\number\month/\number\day/\number\yearltd\ \ \hourmin}}%
\headline={\hfil\draftdate}\writelabels\baselineskip=20pt plus 2pt minus 2pt
 {\count255=\time\divide\count255 by 60 \xdef\hourmin{\number\count255}
  \multiply\count255 by-60\advance\count255 by\time
  \xdef\hourmin{\hourmin:\ifnum\count255<10 0\fi\the\count255}}}
\def\nolabels{\def\wrlabeL##1{}\def\eqlabeL##1{}\def\reflabeL##1{}}
\def\writelabels{\def\wrlabeL##1{\leavevmode\vadjust{\rlap{\smash%
{\line{{\escapechar=` \hfill\rlap{\sevenrm\hskip.03in\string##1}}}}}}}%
\def\eqlabeL##1{{\escapechar-1\rlap{\sevenrm\hskip.05in\string##1}}}%
\def\reflabeL##1{\noexpand\llap{\noexpand\sevenrm\string\string\string##1}}}
\nolabels
%
\global\newcount\secno \global\secno=0
\global\newcount\meqno \global\meqno=1
\def\newsec#1{\global\advance\secno by1\message{(\the\secno. #1)}
\global\subsecno=0\eqnres@t\noindent{\bf\the\secno. #1}
\writetoca{{\secsym} {#1}}\par\nobreak\medskip\nobreak}
\def\eqnres@t{\xdef\secsym{\the\secno.}\global\meqno=1\bigbreak\bigskip}
\def\sequentialequations{\def\eqnres@t{\bigbreak}}\xdef\secsym{}
\global\newcount\subsecno \global\subsecno=0
\def\subsec#1{\global\advance\subsecno by1\message{(\secsym\the\subsecno. #1)}
\ifnum\lastpenalty>9000\else\bigbreak\fi
\noindent{\it\secsym\the\subsecno. #1}\writetoca{\string\quad 
{\secsym\the\subsecno.} {#1}}\par\nobreak\medskip\nobreak}
\def\appendix#1#2{\global\meqno=1\global\subsecno=0\xdef\secsym{\hbox{#1.}}
\bigbreak\bigskip\noindent{\bf Appendix #1. #2}\message{(#1. #2)}
\writetoca{Appendix {#1.} {#2}}\par\nobreak\medskip\nobreak}
%
%
\def\eqnn#1{\xdef #1{(\secsym\the\meqno)}\writedef{#1\leftbracket#1}%
\global\advance\meqno by1\wrlabeL#1}
\def\eqna#1{\xdef #1##1{\hbox{$(\secsym\the\meqno##1)$}}
\writedef{#1\numbersign1\leftbracket#1{\numbersign1}}%
\global\advance\meqno by1\wrlabeL{#1$\{\}$}}
\def\eqn#1#2{\xdef #1{(\secsym\the\meqno)}\writedef{#1\leftbracket#1}%
\global\advance\meqno by1$$#2\eqno#1\eqlabeL#1$$}
%
\newskip\footskip\footskip14pt plus 1pt minus 1pt 
\def\footnotefont{\ninepoint}\def\f@t#1{\footnotefont #1\@foot}
\def\f@@t{\baselineskip\footskip\bgroup\footnotefont\aftergroup\@foot\let\next}
\setbox\strutbox=\hbox{\vrule height9.5pt depth4.5pt width0pt}
\global\newcount\ftno \global\ftno=0
\def\foot{\global\advance\ftno by1\footnote{$^{\the\ftno}$}}
%
\newwrite\ftfile   
\def\footend{\def\foot{\global\advance\ftno by1\chardef\wfile=\ftfile
$^{\the\ftno}$\ifnum\ftno=1\immediate\openout\ftfile=foots.tmp\fi%
\immediate\write\ftfile{\noexpand\smallskip%
\noexpand\item{f\the\ftno:\ }\pctsign}\findarg}%
\def\footatend{\vfill\eject\immediate\closeout\ftfile{\parindent=20pt
\centerline{\bf Footnotes}\nobreak\bigskip\input foots.tmp }}}
\def\footatend{}
%
%
\global\newcount\refno \global\refno=1
\newwrite\rfile
\def\ref{[\the\refno]\nref}
\def\nref#1{\xdef#1{[\the\refno]}\writedef{#1\leftbracket#1}%
\ifnum\refno=1\immediate\openout\rfile=refs.tmp\fi
\global\advance\refno by1\chardef\wfile=\rfile\immediate
\write\rfile{\noexpand\item{#1\ }\reflabeL{#1\hskip.31in}\pctsign}\findarg}
\def\findarg#1#{\begingroup\obeylines\newlinechar=`\^^M\pass@rg}
{\obeylines\gdef\pass@rg#1{\writ@line\relax #1^^M\hbox{}^^M}%
\gdef\writ@line#1^^M{\expandafter\toks0\expandafter{\striprel@x #1}%
\edef\next{\the\toks0}\ifx\next\em@rk\let\next=\endgroup\else\ifx\next\empty%
\else\immediate\write\wfile{\the\toks0}\fi\let\next=\writ@line\fi\next\relax}}
\def\striprel@x#1{} \def\em@rk{\hbox{}} 
\def\lref{\begingroup\obeylines\lr@f}
\def\lr@f#1#2{\gdef#1{\ref#1{#2}}\endgroup\unskip}

\def\addref#1{\immediate\write\rfile{\noexpand\item{}#1}} 
\def\startrefs#1{\immediate\openout\rfile=refs.tmp\refno=#1}
\def\xref{\expandafter\xr@f}\def\xr@f[#1]{#1}
\def\refs#1{\count255=1[\r@fs #1{\hbox{}}]}
\def\r@fs#1{\ifx\und@fined#1\message{reflabel \string#1 is undefined.}%
\nref#1{need to supply reference \string#1.}\fi%
\vphantom{\hphantom{#1}}\edef\next{#1}\ifx\next\em@rk\def\next{}%
\else\ifx\next#1\ifodd\count255\relax\xref#1\count255=0\fi%
\else#1\count255=1\fi\let\next=\r@fs\fi\next}
%

%
\newwrite\ffile\global\newcount\figno \global\figno=1
\def\fig{fig.~\the\figno\nfig}
\def\nfig#1{\xdef#1{fig.~\the\figno}%
\writedef{#1\leftbracket fig.\noexpand~\the\figno}%
\ifnum\figno=1\immediate\openout\ffile=figs.tmp\fi\chardef\wfile=\ffile%
\immediate\write\ffile{\noexpand\medskip\noexpand\item{Fig.\ \the\figno. }
\reflabeL{#1\hskip.55in}\pctsign}\global\advance\figno by1\findarg}
\def\xfig{\expandafter\xf@g}\def\xf@g fig.\penalty\@M\ {}
\def\figs#1{figs.~\f@gs #1{\hbox{}}}
\def\f@gs#1{\edef\next{#1}\ifx\next\em@rk\def\next{}\else
\ifx\next#1\xfig #1\else#1\fi\let\next=\f@gs\fi\next}
\newwrite\lfile
{\escapechar-1\xdef\pctsign{\string\%}\xdef\leftbracket{\string\{}
\xdef\rightbracket{\string\}}\xdef\numbersign{\string\#}}

\def\writestop{\def\writestoppt{\immediate\write\lfile{\string\pageno%
\the\pageno\string\startrefs\leftbracket\the\refno\rightbracket%
\string\def\string\secsym\leftbracket\secsym\rightbracket%
\string\secno\the\secno\string\meqno\the\meqno}\immediate\closeout\lfile}}
\def\writestoppt{}\def\writedef#1{}
\def\seclab#1{\xdef #1{\the\secno}\writedef{#1\leftbracket#1}\wrlabeL{#1=#1}}
\def\subseclab#1{\xdef #1{\secsym\the\subsecno}%
\writedef{#1\leftbracket#1}\wrlabeL{#1=#1}}
\newwrite\tfile \def\writetoca#1{}
\def\leaderfill{\leaders\hbox to 1em{\hss.\hss}\hfill}
\def\writetoc{\immediate\openout\tfile=toc.tmp 
   \def\writetoca##1{{\edef\next{\write\tfile{\noindent ##1 
   \string\leaderfill {\noexpand\number\pageno} \par}}\next}}}

%
\catcode`\@=12 
%
\edef\tfontsize{\ifx\answ\bigans scaled\magstep3\else scaled\magstep4\fi}
\font\titlerm=cmr10 \tfontsize \font\titlerms=cmr7 \tfontsize
\font\titlermss=cmr5 \tfontsize \font\titlei=cmmi10 \tfontsize
\font\titleis=cmmi7 \tfontsize \font\titleiss=cmmi5 \tfontsize
\font\titlesy=cmsy10 \tfontsize \font\titlesys=cmsy7 \tfontsize
\font\titlesyss=cmsy5 \tfontsize \font\titleit=cmti10 \tfontsize
\skewchar\titlei='177 \skewchar\titleis='177 \skewchar\titleiss='177
\skewchar\titlesy='60 \skewchar\titlesys='60 \skewchar\titlesyss='60
\def\titlefont{\def\rm{\fam0\titlerm}
\textfont0=\titlerm \scriptfont0=\titlerms \scriptscriptfont0=\titlermss
\textfont1=\titlei \scriptfont1=\titleis \scriptscriptfont1=\titleiss
\textfont2=\titlesy \scriptfont2=\titlesys \scriptscriptfont2=\titlesyss
\textfont\itfam=\titleit \def\it{\fam\itfam\titleit}\rm}
 \ifx\answ\bigans\else scaled\magstep1\fi
\ifx\answ\bigans\def\abstractfont{\tenpoint}\else
\font\abssl=cmsl10 scaled \magstep1
\font\absrm=cmr10 scaled\magstep1 \font\absrms=cmr7 scaled\magstep1
\font\absrmss=cmr5 scaled\magstep1 \font\absi=cmmi10 scaled\magstep1
\font\absis=cmmi7 scaled\magstep1 \font\absiss=cmmi5 scaled\magstep1
\font\abssy=cmsy10 scaled\magstep1 \font\abssys=cmsy7 scaled\magstep1
\font\abssyss=cmsy5 scaled\magstep1 \font\absbf=cmbx10 scaled\magstep1
\skewchar\absi='177 \skewchar\absis='177 \skewchar\absiss='177
\skewchar\abssy='60 \skewchar\abssys='60 \skewchar\abssyss='60
\def\abstractfont{\def\rm{\fam0\absrm}
\textfont0=\absrm \scriptfont0=\absrms \scriptscriptfont0=\absrmss
\textfont1=\absi \scriptfont1=\absis \scriptscriptfont1=\absiss
\textfont2=\abssy \scriptfont2=\abssys \scriptscriptfont2=\abssyss
\textfont\itfam=\bigit \def\it{\fam\itfam\bigit}\def\footnotefont{\tenpoint}%
\textfont\slfam=\abssl \def\sl{\fam\slfam\abssl}%
\textfont\bffam=\absbf \def\bf{\fam\bffam\absbf}\rm}\fi
\def\tenpoint{\def\rm{\fam0\tenrm}
\textfont0=\tenrm \scriptfont0=\sevenrm \scriptscriptfont0=\fiverm
\textfont1=\teni  \scriptfont1=\seveni  \scriptscriptfont1=\fivei
\textfont2=\tensy \scriptfont2=\sevensy \scriptscriptfont2=\fivesy
\textfont\itfam=\tenit \def\it{\fam\itfam\tenit}\def\footnotefont{\ninepoint}%
\textfont\bffam=\tenbf \def\bf{\fam\bffam\tenbf}\def\sl{\fam\slfam\tensl}\rm}
\font\ninerm=cmr9  \font\ninei=cmmi9 \font\sixi=cmmi6 
\font\ninesy=cmsy9 \font\sixsy=cmsy6 \font\ninebf=cmbx9 
\font\nineit=cmti9 \font\ninesl=cmsl9 \skewchar\ninei='177
\skewchar\sixi='177 \skewchar\ninesy='60 \skewchar\sixsy='60 
\def\ninepoint{\def\rm{\fam0\ninerm}
\textfont0=\ninerm \scriptfont0=\sixrm \scriptscriptfont0=\fiverm
\textfont1=\ninei \scriptfont1=\sixi \scriptscriptfont1=\fivei
\textfont2=\ninesy \scriptfont2=\sixsy \scriptscriptfont2=\fivesy
\textfont\itfam=\ninei \def\it{\fam\itfam\nineit}\def\sl{\fam\slfam\ninesl}%
\textfont\bffam=\ninebf \def\bf{\fam\bffam\ninebf}\rm} 
%
%
\def\noblackbox{\overfullrule=0pt}
\hyphenation{anom-aly anom-alies coun-ter-term coun-ter-terms}
\def\inv{^{\raise.15ex\hbox{${\scriptscriptstyle -}$}\kern-.05em 1}}

\def\Dsl{\,\raise.15ex\hbox{/}\mkern-13.5mu D} 
\def\dsl{\raise.15ex\hbox{/}\kern-.57em\partial}

\catcode`\@=11 

\def\slashit#1{\mathord{\mathpalette\c@ncel{#1}}}
\catcode`\@=12
 \def\Tr{{\rm Tr}}
\font\bigit=cmti10 scaled \magstep1
\def\lspace{\ifx\answ\bigans{}\else\qquad\fi}
\def\lbspace{\ifx\answ\bigans{}\else\hskip-.2in\fi} 
\def\boxeqn#1{\vcenter{\vbox{\hrule\hbox{\vrule\kern3pt\vbox{\kern3pt
	\hbox{${\displaystyle #1}$}\kern3pt}\kern3pt\vrule}\hrule}}}
\def\mbox#1#2{\vcenter{\hrule \hbox{\vrule height#2in
		\kern#1in \vrule} \hrule}}  
%
 \def\CO{{\cal O}} 
    
 \def\CH{{\cal H}}  
  \def\CD{{\cal D}}

\def\darr#1{\raise1.5ex\hbox{$\leftrightarrow$}\mkern-16.5mu #1}

\def\half{{\textstyle{1\over2}}} 
\def\roughly#1{\raise.3ex\hbox{$#1$\kern-.75em\lower1ex\hbox{$\sim$}}}


\def\fonttest{y}
\ifx\boringfonts\fonttest\else
\font\ninerm=cmr9\font\ninei=cmmi9\font\nineit=cmti9\font\ninesy=cmsy9
\font\ninebf=cmbx9\font\ninesl=cmsl9
\def\ninepoint{\def\rm{\fam0\ninerm}
\textfont0=\ninerm \scriptfont0=\sevenrm \scriptscriptfont0=\fiverm
\textfont1=\ninei  \scriptfont1=\seveni  \scriptscriptfont1=\fivei
\textfont2=\ninesy \scriptfont2=\sevensy \scriptscriptfont2=\fivesy
\textfont\itfam=\nineit \def\it{\fam\itfam\nineit} \def\sl{\fam\slfam\ninesl}
\textfont\bffam=\ninebf \def\bf{\fam\bffam\ninebf} \rm}
\fi

\hyphenation{anom-aly anom-alies coun-ter-term coun-ter-terms
dif-feo-mor-phism dif-fer-en-tial super-dif-fer-en-tial dif-fer-en-tials
super-dif-fer-en-tials reparam-etrize param-etrize reparam-etriza-tion}
 
 
%
%
%
\newwrite\tocfile\global\newcount\tocno\global\tocno=1
\ifx\bigans\answ \def\tocline#1{\hbox to 320pt{\hbox to 45pt{}#1}}
\else\def\tocline#1{\line{#1}}\fi
\def\toclead{\leaders\hbox to 1em{\hss.\hss}\hfill}
\def\tnewsec#1#2{\xdef #1{\the\secno}\newsec{#2}
\ifnum\tocno=1\immediate\openout\tocfile=toc.tmp\fi\global\advance\tocno
by1
{\let\the=0\edef\next{\write\tocfile{\medskip\tocline{\secsym\ #2\toclead\the
\count0}\smallskip}}\next}
}
\def\tnewsubsec#1#2{\xdef #1{\the\secno.\the\subsecno}\subsec{#2}
\ifnum\tocno=1\immediate\openout\tocfile=toc.tmp\fi\global\advance\tocno
by1
{\let\the=0\edef\next{\write\tocfile{\tocline{ \ \secsym\the\subsecno\
#2\toclead\the\count0}}}\next}
}
\def\tappendix#1#2#3{\xdef #1{#2.}\appendix{#2}{#3}
\ifnum\tocno=1\immediate\openout\tocfile=toc.tmp\fi\global\advance\tocno
by1
{\let\the=0\edef\next{\write\tocfile{\tocline{ \ #2.
#3\toclead\the\count0}}}\next}
}
%
%

%
%
%
%

%
\long\def\optional#1{}
\def\cmp#1{#1}         
%
%
\let\narrowequiv=\equiv
\def\equiv{\;\narrowequiv\;}

\def\tilde{\widetilde}
\fontdimen16\tensy=2.7pt\fontdimen17\tensy=2.7pt 
 
 
 
%

\def\la{\lambda}
\def\ep{\epsilon}
\def\al{{\alpha}}

\def\dl{\delta}
%
%

\def\CJ{{\cal J}}

\def\CO{{\cal O}}

\def\CH{{\cal H}}

\def\CD{{\cal D}}

%
%
%
\def\boxit#1#2{
        $$\vcenter{\vbox{\hrule\hbox{\vrule\kern3pt\vbox{\kern3pt
	\hbox to #1truein{\hsize=#1truein\vbox{#2}}\kern3pt}\kern3pt\vrule}
        \hrule}}$$
}


%

\def\lfr#1#2{{\textstyle{#1\over#2}}} 



\def\splitexact#1#2{\mathrel{\mathop{\null{
\lower4pt\hbox{$\rightarrow$}\atop\raise4pt\hbox{$\leftarrow$}}}\limits
^{#1}_{#2}}}

%
%
\def\pa{\partial}

\def\pd#1#2{{\partial #1\over\partial #2}} 
%
%
\def\ab{{\bar{\vphantom\i a}}}  
%
%

\def\ex#1{{\rm e}^{#1}}                 
\def\dd{\mskip 1.3mu{\rm d}\mskip .7mu} 


\def\det{\hbox{det$\,$}}                

%
%

\def\IM{isomorphism}

\def\eg{{\it e.g.}}\def\ie{{\it i.e.}}

%
%

\ifx\boringfonts\fonttest
\font\blackboard=cmssbx10 \font\blackboards=cmssbx10 at 7pt  
\font\blackboardss=cmssbx10 at 5pt
\else 
\font\blackboard=msym10 \font\blackboards=msym7   
\font\blackboardss=msym5
\fi
\newfam\black
\textfont\black=\blackboard
\scriptfont\black=\blackboards
\scriptscriptfont\black=\blackboardss

                                  
%
\ifx\boringfonts\fonttest
\font\gothic=cmssbx10 \font\gothics=cmssbx10 at 7pt  
\font\gothicss=cmssbx10 at 5pt
\else
\font\gothic=eufm10 \font\gothics=eufm7
\font\gothicss=eufm5
\fi
\newfam\gothi
\textfont\gothi=\gothic
\scriptfont\gothi=\gothics
\scriptscriptfont\gothi=\gothicss

{\catcode`\@=11\gdef\oldcal{\fam\tw@}}
\newfam\curly
\ifx\boringfonts\fonttest\else
\font\curlyfont=eusm10 \font\curlyfonts=eusm7
\font\curlyfontss=eusm5
\textfont\curly=\curlyfont
\scriptfont\curly=\curlyfonts
\scriptscriptfont\curly=\curlyfontss
\def\cal{\fam\curly\relax}    
\fi
%

\ifx\boringfonts\fonttest\else\fi

\global\newcount\pnfigno \global\pnfigno=1
\newwrite\ffile
\def\pfig#1#2{Fig.~\the\pnfigno\pnfig#1{#2}}
\def\pnfig#1#2{\xdef#1{Fig. \the\pnfigno}%
\ifnum\pnfigno=1\immediate\openout\ffile=figs.tmp\fi%
\immediate\write\ffile{\noexpand\item{\noexpand#1\ }#2}%
\global\advance\pnfigno by1}

%
%
\def\tfig#1{Fig.~\the\pnfigno\xdef#1{Fig.~\the\pnfigno}\global\advance\pnfigno by1}

%
%
%
%
\def\figI{y}
\def\ifigure#1#2#3#4{
\midinsert
\ifx\figflag\figI
\vbox to #4truein{
\vfil\centerline{\epsfysize=#4truein\epsfbox{#3}}}
\else
\vbox to .2truein{}
\fi
\narrower\narrower\noindent{\bf #1:} #2
\endinsert
}








%
%

%


\def\inbar{\,\vrule height1.5ex width.4pt depth0pt}
\def\IB{\relax{\rm I\kern-.18em B}}
\def\IC{\relax\hbox{$\inbar\kern-.3em{\rm C}$}}
\def\ID{\relax{\rm I\kern-.18em D}}
\def\IE{\relax{\rm I\kern-.18em E}}
\def\IF{\relax{\rm I\kern-.18em F}}
\def\IG{\relax\hbox{$\inbar\kern-.3em{\rm G}$}}
\def\IH{\relax{\rm I\kern-.18em H}}
\def\II{\relax{\rm I\kern-.18em I}}
\def\IK{\relax{\rm I\kern-.18em K}}
\def\IL{\relax{\rm I\kern-.18em L}}
\def\IM{\relax{\rm I\kern-.18em M}}
\def\IN{\relax{\rm I\kern-.18em N}}
\def\IO{\relax\hbox{$\inbar\kern-.3em{\rm O}$}}
\def\IP{\relax{\rm I\kern-.18em P}}
\def\IQ{\relax\hbox{$\inbar\kern-.3em{\rm Q}$}}
\def\IR{\relax{\rm I\kern-.18em R}}
\font\cmss=cmss10 \font\cmsss=cmss10 at 10truept
\def\IZ{\relax\ifmmode\mathchoice
{\hbox{\cmss Z\kern-.4em Z}}{\hbox{\cmss Z\kern-.4em Z}}
{\lower.9pt\hbox{\cmsss Z\kern-.36em Z}}
{\lower1.2pt\hbox{\cmsss Z\kern-.36em Z}}\else{\cmss Z\kern-.4em Z}\fi}
\def\IGa{\relax\hbox{${\rm I}\kern-.18em\Gamma$}}
\def\IPi{\relax\hbox{${\rm I}\kern-.18em\Pi$}}
\def\ITh{\relax\hbox{$\inbar\kern-.3em\Theta$}}
\def\IOm{\relax\hbox{$\inbar\kern-3.00pt\Omega$}}

\def\pndate{1/94}
\long\def\suppress#1{}
\suppress{\def\boringfonts{y}  
\def\pndate{\vbox{\vskip .2truein\hbox{{\sl Running title:} Measure Factors and
Correlations of Membranes}\hbox{{\sl PACS:} 87.10, 11.17, 68.15, 87.20}
}}
\def\figflag{n}
\baselineskip=20pt
\def\ifigure#1#2#3#4{\nfig\dumfig{#2}}

} 

\let\epsilon=\varepsilon
\ifx\answ\bigans \else\noblackbox\fi

\Title{\vbox{\hbox{UPR--599T}}}{\vbox{\centerline{Measure Factors, Tension, and
}
\vskip2pt\centerline{Correlations of Fluid Membranes}}}

\centerline{W. Cai, T.C. Lubensky, P. Nelson and T. Powers}\smallskip
\centerline{Physics Department, University of Pennsylvania}
\centerline{Philadelphia, PA 19104 USA}
\bigskip\bigskip

We study two geometrical factors needed for the correct construction
of statistical ensembles of surfaces. Such ensembles appear in the
study of fluid bilayer membranes, though our results are more
generally applicable. The naive functional measure over height
fluctuations must be corrected by these factors in order to give
correct, self-consistent formulas for the free energy and correlation
functions of the height. While one of these corrections --- the
Faddeev-Popov determinant --- has been studied extensively, our
derivation proceeds from very simple geometrical ideas, which we hope
removes some of its mystery. The other factor is similar to the
Liouville correction in string theory. Since our formulas differ from
those of previous authors, we include some explicit calculations of
the effective frame tension and two-point function to show that
our version indeed secures coordinate-invariance and consistency to
lowest nontrivial order in a temperature expansion.

\Date{\pndate}\noblackbox                 

\def\roughness{$q^2$ }\def\ei{eigen}

\def\ev{{\bf e}}\def\Rv{{\bf R}}\def\nv{{\bf n}}

\def\Ga{\Gamma}\def\Gah{\Gamma[\bar h]}\def\La{\Lambda}
\def\fpd#1#2{{\delta #1\over\delta #2}} 
\def\fpdl#1#2#3{\left.{\delta #1\over\delta #2}\right|_{#3}} 
\def\ab{{A_B}}\def\iab{} 
\def\ij{_{ij}}\def\gb{{\bar g}}
\def\dh{[Dh]}
\def\wgt{\ex{-\CH/T}}
\def\hb{\bar h}\def\Db{\bar\Delta}\def\trg{\Tr_\gb\log}
\def\du{\dd^2u\,}\def\rgb{\sqrt{\bar g}}\def\duu{\dd^2u'\,}
\def\dk{{\dd^2k\over (2\pi)^2}\,}
\def\dq{{\dd^2q\over (2\pi)^2}\,}\def\dqq{{\dd^2q'\over (2\pi)^2}\,}
\def\dhs{(\pa h)^2}\def\dhbs{(\pa \bar h)^2}
\def\FP{_{\rm FP}}\def\rfp{r\FP}\def\fp{Faddeev-Popov}
\def\ko{\kappa_0}\def\muo{\mu_0}\def\mul{\mu_1}\def\dlh{\dl\CH}
\def\go{{g_0}}\def\goab{g_{0,ab}}\def\rgo{\sqrt{g_0}}\def\trgo{\Tr_\go\log}
\def\eff{{\rm eff}}  \def\gzer{$\Ga^{(0)}$}
\def\naive{_{\rm naive}}
\def\la{\langle}
\def\ra{\rangle}
\def\be{{\bf e}}
\def\bn{{\bf n}}
\def\bR{{\bf R}}
\def\part{\partial}
\def\vol{{\rm vol}}

\def\intprm{\int^{\prime}}
\long\def\fixup#1{#1}
\hfuzz=3truept

\lref\Cates{
M. Cates, ``The Liouville field theory of random surfaces,'' Europhys.
Lett. {\bf8} (1988) 719.\optional{[6/88]}}%
\lref\DeTa{P.G. de\thinspace Gennes and C.  Taupin, J. Phys. Chem.
{\bf86} (1982) 2294.}
\def\kruns{\DeTa--%
\nref\Helfa{W. Helfrich, J. Phys. (Paris) {\bf46} (1985) 1263; {\it
ibid.} {\bf47} (1986) 321.}%
\nref\PeLe{L. Peliti and S. Leibler, ``Effects of thermal fluctuations on
systems with small surface tension,''
Phys. Rev. Lett. {\bf54} (1985) 1690.\optional{[2/4/85]}}%
\nref\Fors{D. F\"orster, ``On the scale dependence, due to thermal
fluctuations, of the elastic properties of membranes,''
Phys. Lett {\bf114A} (1986) 115.\optional{[9/85]}}%
\nref\Pold{A. Polyakov, ``Fine structure of strings,''
Nucl. Phys. {\bf B268} (1986) 406.}
\Kla}

\lref\Messager{R. Messager, P. Bassereau, and G. Porte, ``Dynamics of
the undulation mode in swollen lamellar phases,'' J. Phys. France
{\bf51} (1990) 1329.}
\lref\bigchiral{P. Nelson and T. Powers, ``Renormalization of chiral
couplings in tilted bilayer membranes,'' J.  Phys. France II  {\bf3}
(1993) 1535.}
\def\muruns{(for example, see \PeLe%
\Fors%
\Kla
--\nref\Klb{H. Kleinert, ``The membrane properties of condensing
strings,'' Phys. Lett. {\bf174B} (1986) 335.\optional{[12/85]}}%
\nref\Meunier{J. Meunier, J. Phys. (France) {\bf48} (1987) 1819.
\optional{[3/87]}}%
\nref\DSMMS{F. David, in \SMMS.}%
\DaLe)}
\lref\tclwc{W. Cai and T.C. Lubensky, ``Covariant hydrodynamics of
fluid membranes'', preprint (1993).}
\lref\MiMo{D. Morse and S. Milner, ``Fluctuations and phase behavior of
surfactant vesicles,'' Preprint 1993.}
\lref\CaHe{P. Canham , J. Theor. Biol. {\bf26} (1970) 61; W.
Helfrich, Naturforsch. {\bf28C} (1973) 693.}
\lref\SMMS{See for example {\sl Statistical mechanics of membranes and
surfaces,} D. Nelson {\it et al.}, eds (World Scientific, 1989).}
\lref\DaLe{F. David and S. Leibler, ``Vanishing tension of fluctuating
membranes,'' J. Phys. II France {\bf 1} (1991) 959.} 
\lref\Kla{H. Kleinert, ``Thermal softening of curvature elasticity in
membranes,''
Phys. Lett. {\bf 114A} (1986) 263.\optional{[11/14/85]}}%
\lref\Can{P. Canham , J. Theor. Biol. {\bf26} (1970) 61}
\lref\Helfaa{W.
Helfrich, Naturforsch. {\bf28C} (1973) 693\optional{[introduces tilt
by analogy with lyotropic l.c.'s!]}.}
\lref\PNTRP{P. Nelson and T. Powers, Phys. Rev. Lett. {\bf69} (1992) 3409.}
\lref\MFM{G. Moore and P. Nelson, \cmp{``Measure for moduli,''} Nucl. Phys.
{\bf B266} (1986) 58.}
\lref\Polch{J. Polchinski, \cmp{``Evaluation of the one loop string path
integral,''} Commun. Math. Phys. {\bf 104} (1986) 37.}
\lref\Browicz{E. Browicz, Zbl. Med. Wiss. {\bf28} (1890) 625.}
\lref\DDK{F. David, ``Conformal field theories coupled to 2-D gravity in the
conformal
gauge,''
 Mod. Phys. Lett. {\bf A3} (1988) 1651; J. Distler and H. Kawai,
``Conformal field theory and 2-D quantum gravity,''  Nucl. Phys. {\bf
B321} (1989) 509.}
\lref\PoStro{J. Polchinski and A. Strominger, ``Effective string theory,''
Phys. Rev. Lett. {\bf67} (1991) 1681.}
\lref\Polbook{A. Polyakov, {\sl Gauge fields and strings}, (Harwood,
1987).}


\newsec{Introduction and Summary}
Amphiphilic molecules in water tend to aggregate into thin flexible bilayers,
which in turn form a wide variety of nearly two-dimensional structures with
characteristic size of order microns \SMMS. The study of these structures
may become significant for biology: they are certainly important
technologically as the underlying elements of microemulsions and other
complex fluids.

To understand the static and dynamic properties of bilayer membranes, and
the transitions between different morphologies, we must first understand
the role of thermal fluctuations in two-dimensional surfaces, a
subtle problem in statistical mechanics. Thermal fluctuations are
important because regardless how stiff a membrane may be, very long-wavelength
undulations in its shape are always allowed; on long enough length scales
any membrane will appear flexible and will undulate significantly, a phenomenon
first described in the cell walls of red blood cells in the 19th century
\Browicz.\foot{Strictly speaking we are here talking about ``fluid membranes,''
those with no internal structure or order within the surface. In this paper
we discuss only fluid membranes, but our measure factors can be used in
more complicated situations too \bigchiral.}

Since the length scales we wish to study are much larger than the size of
the constituent molecules, we expect that a membrane will be characterized
by just a few effective parameters, summarizing the effects of the complicated
molecular forces. Indeed, the famous Canham-Helfrich model\CaHe\ describes the
equilibrium statistical mechanics of membranes in terms of just two parameters:
a chemical potential for the addition of surfactant molecules, which
we will call $\mu_0$, and a bending energy
coefficient, the stiffness $\kappa_0$. (Another parameter, the
gaussian stiffness $\bar\kappa_0$, only enters when we consider
topology change.)
The bare coefficients $\mu_0,\kappa_0$ are not directly
observable, but from them we can derive some more relevant
phenomenological parameters.
We will mainly study the ``frame tension''
$\tau$, which is the free energy per unit area of the fluctuating
membrane, and the two-point correlation
$\la h(u)h(0)\ra$ of the height $h(u)$ of the membrane from its average
plane, which we fix by fixing the edges of the membrane to lie on a
square frame. The leading behavior of this two-point function at small
wavenumber
defines another phenomenological parameter, the ``\roughness  coefficient''
$r$ via
\eqn\esigd{
\la h(q)h(-q)\ra={A_BT\over
rq^2+\CO(q^4)}\quad,}
where $A_B$ is the area of the frame, $T$ is the temperature
(we set Boltzmann's constant $k_B$ equal to one), and $q$
is a wavenumber.
Another reason to introduce $r$ is that in dynamics problems we can
still calculate correlation functions like \esigd, while there is no
analog of the free energy from which to compute  $\tau$.
Two of us study the dynamical problem
in \tclwc. 

In this paper we will explore the relation between the various coefficients
$\mu_0$, $\kappa_0$, $\tau$, and $r$. More generally, on long length scales
the precise size of the constituent molecules is immaterial: we can get the
same answers using molecules of size $a$ with coefficients $\mu_0$, $\kappa_0$
or with rescaled molecules of size $b^{-1}a$, $b<1$, and effective coefficients
$\mu_\eff(b)$, $\kappa_\eff(b)$. The scale dependence of $\kappa_\eff$ is well
known \kruns, but various answers for $\mu_\eff$ have been given \muruns.
In part the differences reflect convention, but there is a key physics point
which we will address: to get  correct answers we must define our
functional measure properly. Two of us have already discussed such
matters in \bigchiral, but here we will add a number of points, as follows.

We first introduce the effective action $\Gah$ of the fluctuating
membrane and recall the argument of \bigchiral\ that its form is
constrained by the coordinate invariance of the underlying theory. In
particular we show in sect.~2 that the coefficients $r$ and $\tau$
defined above must be equal. We then set up a naive, uncorrected,
calculational scheme which yields (consistent with \DaLe)
\eqn\etaur{
\tau=\mu_0+{T\over2}\int^\Lambda {\dd^2q\over(2\pi)^2}\,
\log\left[(\mu_0q^2+\kappa_0q^4)\,
{a^2\lambda^2\over 2\pi T}\right]\quad,}
where $\Lambda=2\pi/a$ and $a$ is the linear size of the constituent
molecules. Following Morse
and Milner we have introduced the thermal de~Broglie wavelength
\eqn\elamb{\lambda=
h/\sqrt{2\pi mT}\quad,}
 where $m$ is the mass of a surfactant molecule \MiMo.
The same calculational scheme however yields a very different formula
for $r$. In fact \etaur\ is our final result, but clearly we need to
work a bit harder to understand $r$.

The first correction factor, the ``Faddeev-Popov'' factor, is well
known. In \bigchiral\ we gave a formula for this factor which differs
from the one given by David \DSMMS; here in sect.~3 we give a simple
geometrical motivation followed by a derivation of our formula which
is simpler than the one in \bigchiral. While unimportant in the
calculation of $\tau$, this factor does modify the naive calculation
of $r$ (and higher correlation functions too), yielding an $\rfp$
which still disagrees with $\tau$.

Next in sect.~4 we argue that another correction factor is needed.
Physically our membrane consists of constituent molecules each
occupying a fixed area $a^2$ in physical 3-space. As the surface bows
outward away from its flat equilibrium configuration its area
increases and additional molecules must be brought in from solution,
effectively increasing the number of degrees of freedom in the
membrane problem. A statistical measure with variable number of
degrees of freedom, depending on the configuration itself, is rather
complicated. We would prefer a measure with fixed number of degrees of
freedom, as  indeed we used implicitly in the naive derivation of
$\rfp$. Since changing the number of degrees of freedom, or
equivalently the ultraviolet cutoff, can be compensated by
renormalizing the bare energy, we expect that a {\it counterterm}
correction will need to be added to the naive derivation.

We fix this counterterm by requiring consistency between two different
Monge-gauge calculations of the free energy (in the Appendix we give a
different strategy). \optional{
 In this way we
discover that
\eqn\ertau{r=\tau=\rfp + \left.\pd{\tau}{\log \La^1}\right|_{\muo,\ko}\quad,}
where we introduced an anisotropic wavenumber
cutoff $(\La^1,\La^2)$. In fact,
substituting our formulas for $\tau$ and $\rfp$ into \ertau\ shows
that they indeed obey \ertau.}
The required counterterm turns out to make a contribution precisely
changing the $q^2$ coefficient from $r\FP$ to $r=\tau$.

The introduction of our second correction factor may seem like ad hoc
wish-fulfillment, but in the Appendix we give some detailed
calculations showing how it and the \fp\ factor together are crucial
to obtain consistent answers; in particular we show that the specific
coefficient of the counterterm implied by consistency is the right
one to get $r=\tau$.
Furthermore, an analogous correction is well known in the string theory
literature, where it is called the ``Liouville factor'' \DDK\PoStro.
This factor sometimes appears in the previous membrane literature
(\eg\ \Cates), but one gets
the impression that it matters only in the low-stiffness regime beyond
the persistence length; again, we will see that it is needed to get
the correct correlation functions even in the more physical stiff
regime. Our correction differs in detail from the Liouville factor
because the latter is appropriate for conformal gauge, but in each
case the motivation is the same.
We will also argue that the usual interpretation of
the Liouville factor as a Jacobian is misleading; really as argued
above it reflects renormalization.

Again our counterterm is important because without it, ordinary
diagrammatic perturbation theory gives incorrect results due to the
subtlety in the statistical measure discussed above.
Finally we conclude in sect.~5.

\newsec{General Arguments}

We will consider fluid membranes, those with no internal order or structure.
Real membranes are typically fluid at temperatures high enough to destroy
in-plane order, but in any case such order can easily be incorporated into
the argument below \bigchiral. The constituents of a membrane have a
preferred spacing, with a high energy cost for local deviations from that
density. Thus, in this regime the free energy cost of a membrane configuration
depends only on its {\it shape}.

As mentioned in the Introduction, we can thus capture the physics of length
scales much longer than the constituent molecule size by regarding the
surface as continuous and using the
Canham-Helfrich free energy \CaHe
\eqn\eCaHe{
\CH=\mu_0\int\dd S+{\kappa_0\over2}
\int\dd S\left({1\over R_1}+{1\over R_2}\right)^2\quad.}
Here $\dd S$ is the element of surface area, $\int\dd S$ is the total area
of the surface, and $R_1$, $R_2$ are the principal radii of curvature of the
surface. The two coefficients $\mu_0$, $\kappa_0$ are respectively the area
cost and bending rigidity.

The form of \eCaHe\ is severely constrained by the requirement that the
free energy depend only on the membrane's shape, and not on how one
chooses to label points on the surface. While this property is manifest
in \eCaHe, to do calculations a more explicit version of this equation
proves necessary, in which we do choose coordinates $u=(u^1,u^2)$ to
label points $\bR(u)$ of the surface. As usual, we then get tangent vectors
$\be_a=\part_a\bR\equiv\part\bR/\part u^a$ at $u$ and a metric tensor
$g_{ab}=\be_a\cdot\be_b$ at each point. The unit normal is then
$\bn=(\be_1\times\be_2)/|\be_1\times\be_2|$, and the curvature tensor is
$K_{ab}=\bn\cdot\nabla_a\part_b\bR$, where $\nabla_a$ is the covariant
derivative associated to the metric $g_{ab}$. Letting $g\equiv\det[g_{ab}]$,
$[g^{ab}]=[g_{ab}]^{-1}$, and $K_a^a\equiv g^{ab}K_{ab}$, eqn.~\eCaHe
\ becomes
\eqn\eCaHeb{
\CH=\mu_0\int d^2u\sqrt{g}+{\kappa_0\over2}\int d^2u\sqrt{g}\,(K_a^a)^2\quad.}

If we change the coordinates from $u^a$ to $\tilde u^a=\tilde u^a(u)$ we just
relabel points in the plane. For an infinitesimal change $\tilde u^a=
u^a+\epsilon^a(u)$ we get a new $\tilde\bR(\tilde u)=\bR(u)$ describing
the same surface:
\eqn\ecoord{
\tilde\bR(u)=\bR(u)-\ep^a(u)\be_a(u)\quad.}
The change of $\bR$ is a {\it tangential} motion; conversely, tangential
changes of $\bR$ are not physical changes to the shape of the surface and
ought not to be counted separately in the statistical sum. We thus need
to supplement \eCaHeb\ by choosing a protocol for assigning a single
coordinate system to each shape, and a procedure for discarding every
surface $\bR(u)$ not parameterized in this way. Any such choice is called
a ``gauge-fixing'' procedure.

A popular class of gauge-fixing procedures are the so-called ``normal gauges.''
Suppose we have arranged that configurations very close to one reference
surface will dominate the statistical sum (by stretching the membrane across
a fixed frame, for instance). We choose once and for all a parameterization
$\bR_0(u)$ for this reference surface and compute its unit normal $\bn_0(u)$.
Then other nearby surfaces can be written as
\eqn\engauge{
\bR(u)=\bR_0(u)+h(u)\bn_0(u)}
where the height field $h$ at any point is uniquely defined as the normal
distance from that point to the reference surface. We thus describe each
distinct surface once when we sum over all functions $h(u)$.
Of course the point of this paper is that the correct definition of
this sum is a tricky business, but let us proceed formally to obtain
some general properties constraining the correct prescription.

Let us consider a membrane confined to span a square, flat frame of
area $A_B$. We suppose surfactant molecules can join or leave the
membrane with a fixed cost in free energy of $\mu_0$ per added area
(one can easily pass to an ensemble with a fixed number of molecules by
a Legendre transformation). The full free energy of our system $Z(A_B)$ then
depends on the base area, and implicitly on $\mu_0$, the stiffness
$\kappa_0$ appearing in \eCaHe, the size $a\equiv2\pi/\Lambda$ of the
constituent molecules, and the temperature $T$.

Since our frame is flat, the equilibrium (zero-temperature) surface will
be flat too, and so it is convenient to work in Monge gauge, a normal
gauge with $\bR_0(u)$ the surface spanned by the frame, with Cartesian
coordinates 
 $u^a$ running from zero to $\sqrt{\ab}$. In addition to
$Z(A_B)$ we will find it convenient to introduce the ``effective action''
$\Gamma[\bar h]$. To define $\Gamma[\bar h]$ we introduce a forcing
term into
$\CH$, $\CH_j=\CH-\int\bar hj$, and adjust $j$ to ensure that
$\la h(u)\ra$ is the desired function $\bar h$.\foot{This is a formal
trick to obtain correlations of height; we
do not imagine $j$ as coming from any physical force. Mathematically
$j=j_{12}\dd^2u$ is a density on parameter space, so that $\dl/\dl j(u)$ is a
scalar. For more details see \bigchiral.}  Computing the partition function
\eqn\edZ{Z[A_B,j]=\int\dh \wgt\ex{(1/T)\int h j}\quad,}
 we then let
\eqn\eleg{
\Gamma[\hb]\equiv-T\log Z(A_B,j)+\int \bar hj\quad.}
The
frame tension is the free energy per area of the unforced membrane, so we have
\eqn\edtau{
\tau=-{T\over A_B}\,\log Z(A_B)=
{1\over A_B}\,\Gamma(\hb=0)\equiv
{1\over A_B}\,\Gamma^{(0)}\quad.}

Physically there is nothing special about Monge gauge. Even if we
agree to work in Monge gauge there is nothing special about a
reference surface $\bR_0(u)$ coinciding with the plane of the frame; a
tilted reference surface must give the same free energy. All that
matters is the physical {\it area} of the frame. Thus the terms of
$\Gah$ involving at most first derivatives of $\hb$ (higher
derivatives are insensitive to tilting) must enter in the combination
$\int\du\rgb$, where
$$\bar g_{ab} = \iab\dl_{ab} + \pa_a\hb\pa_b\hb$$
is the induced metric and $\bar g=\det\bar g_{ab}$ as usual. In other
words,
\eqn\elowh{\Gah=\tau\int\du\bigl(\iab1 + \half\dhbs
+\cdots\bigr)+\cdots\quad,}
where the first ellipsis involves $\CO(\hb^4)$ while the second
involves more derivatives than $\hb$'s. We gave a more detailed argument for
this conclusion in \bigchiral.

Whatever the precise form of the measure $\dh$ appearing in \edZ, it
does not depend on $j$, and so the first moment
\eqn\edummy{\la h(u)\ra_j\equiv \int\dh\wgt\ex{(1/T)\int h j}h(u)=TZ\inv
{\dl Z[j]\over\dl j(u)}\quad.}
Differentiating \eleg\ and using $\la h \ra_j=\hb$, we find
\eqn\edummy{{\dl\Ga\over\dl\hb(u)}=j(u)-TZ\inv
\int{\dl j(u')\over\dl \hb(u)} {\dl Z\over\dl j(u')} +
\int {\dl j(u')\over\dl \hb(u)} \hb(u') = j(u)\quad.}
Differentiating again and using $\la\hb\ra_{j=0}=0$, we see that
\eqn\eggg{\eqalign{
\la h(u)h(u')\ra_{j=0}&=
Z\inv T^2\left.\fpd{^2Z}{j(u)\dl j(u')}\right|_0=
Z\inv T\fpdl{}{j(u)}{0}\bigl(Z\hb(u')\bigr)\cr
&=T\fpdl{\hb(u')}{j(u)}{0}=T\left[\fpdl{j(u)}{\hb(u')}{0}\right]\inv
=T\left[\fpdl{^2\Ga}{\hb(u)\dl\hb(u')}{0}\right]\inv
\quad.\cr}
}
Thus the inverse of the two-point function is just the quadratic part
of the effective action.

We have repeated the above well-known argument to make a point: It in
no way depended upon the niceties of the measure $\dh$. Any exotic
gauge-fixing or cutoff-stretching factor which we may eventually fold
into $\dh$ will not alter the above conclusion, though they can and
will affect the relation of the quadratic part 
to the constant part \gzer. Combining \eggg\ with
\elowh, \esigd\ at once shows that the \roughness  coefficient $r$
equals 
 the frame tension $\tau$.

Let us make a first attempt at calculating $r$ and $\tau$. If the
temperature $T$ is small we can expand everything around the
equilibrium configuration $h=0$. The energy functional \eCaHeb\ then
becomes \Kla
\eqn\eHexpa{\eqalign{
\CH=\iab\muo&\int\du\bigl(1+\half\dhs
-\lfr18 (\pa h)^4
\bigr) \cr
+{\ko\over 2\iab}&\int\du\bigl[(\partial^2 h)^2
-\lfr1{2\iab}\dhs(\partial^2 h)^2
-2
(\partial^2 h)\pa_ah\pa_bh\pa_a\pa_bh\bigr]+\CO(h^6)\quad.\cr
} }
In this formula $\pa_a=\pa/\pa u^a$
, $\partial^2=\dl^{ab}
\pa_a\pa_b$ is the flat laplacian, and all indices are contracted with
the 
Kronecker symbol. Thus all $h$-dependence is
explicitly in view in \eHexpa. To calculate $Z(\ab)$ we need a
proposal for $\dh$. The simplest choice is to choose a mesh of grid
points $\{u_{ij}\}$ on $u$ space, spaced by $a
$, and let
\eqn\emnaive{[Dh]\naive =\prod_{ij}(\dd h(u_{ij})/\lambda)\quad.}
We introduced
the thermal wavelength $\lambda$ here to render the measure properly
dimensionless.\foot{In \DaLe\ and \bigchiral\ this scale was taken to
be the molecule size $a$ or a constant multiple of it. This is
legitimate when we do not care about the bare chemical potential, for
example when we just want to dial $\muo$ to zero physical tension. But
$\muo$ does have physical meaning, and
Milner and Morse have argued that the factor in \emnaive\ is the
correct choice \MiMo.}
 We will argue later
that $\dh\naive$ is wrong, but for now we take it as
our starting point.

For low temperature we approximate $Z(\ab)$ as $\ex{-\muo\ab /T}$
times a gaussian integral. Collecting the quadratic terms of \eHexpa\
and writing $\int\du=a^2
\sum_{ij}$ we find
$$\Ga^{(0)}\equiv -T\log Z(\ab)= \muo\ab + {T\over2}\sum \log
\left[{\lambda^2a^2\over 2\pi T\iab}\bigl(\muo q^2+\ko
q^4\bigr)\right]
\quad.$$
Now the sum is over the points $q$ of reciprocal space. 
The upper limit of this sum is related to
the constituent size $a$; for example, if in \emnaive\ we choose
points on a square grid, then our Brillouin zone is a square of length
$2\pi/a$.  
Thus  we recover the announced formula \etaur.

Next we turn to the two-point function, which we provisionally define as
\eqn\edummy{\la h(u)h(0)\ra\naive= Z\inv\int\dh\naive\wgt h(x)h(0)\quad.}
This time however the constant term of \eHexpa\ drops out, the
quadratic term gives the leading contribution $\la h(q)h(-q)\ra_0=
{T\ab\over\muo q^2+\ko q^4}$, and the quartic terms yield the desired
thermal correction to $\la hh\ra_0$ via Wick's theorem. To shorten the
formulas, in the rest of this paragraph we will drop all factors of
$A_B$, or set $A_B=1$. Retaining only those terms contributing to $r$
we get
$$\eqalign{
\la h(u)h(0)\ra\naive =& \la h(u)h(0)\ra_0\cr &+
{\muo\over 8T}\Biggl\{4\int\duu
\la h(u)\pa_a h(u')\ra_0 \la h(0)\pa_a h(u')\ra_0 \la \pa_bh(u')\pa_bh(u')\ra_0
\cr
&\quad+8\int\duu
\la h(u)\pa_a h(u')\ra_0 \la h(0)\pa_b h(u')\ra_0 \la \pa_ah(u')\pa_bh(u')\ra_0
\Biggr\}\cr
&+{\ko\over 4T}\,2\int\duu
\la h(u)\pa_a h(u')\ra_0 \la h(0)\pa_a h(u')\ra_0 \la \partial^2
h(u')\partial^2 h(u')\ra_0\cr
&+{\ko\over T}\,2\int\duu
\la h(u)\pa_a h(u')\ra_0 \la h(0)\pa_b h(u')\ra_0 \la
\pa_a\pa_bh(u')\partial^2 h(u')\ra_0
\ ,\cr}$$
or
$$\eqalign{
\la h(q)h(-q)\ra\naive =& {T\over\ko q^4+\muo q^2}\left[1+{T\over\ko q^4+\muo
q^2}
\int\dqq{\lfr32\ko(q')^4+\muo(q')^2 \over \ko(q')^4+\muo(q')^2 }\right]\cr
=&T\left\{\muo q^2 - {Tq^2\over2}\int\dqq\left(3-{\muo(q')^2\over
\ko(q')^4+\muo(q')^2 }\right)+\CO(q^4)\right\}\inv
\ ,\cr}$$
so
\eqn\ernaive{r\naive=\muo-{T\over2}\int^\Lambda\dq\left(3-{\muo q^2\over
\ko q^4+\muo q^2 }\right)\quad.}
Eqn.~\ernaive\ is essentially the result of Meunier~\Meunier. It
certainly does not equal \etaur, notwithstanding our
formal argument that $r=\tau$.

\newsec{Gauge Fixing}
Apparently our troubles stem from \emnaive. How should we sum over all
configurations? We will return in the next section to the question of
the spacing of grid points, but
even with this choice
made correctly, eqn.~\emnaive\ is not correct. Consider the two surfaces $\bR$,
$\bR+\dd\bR$ in \tfig\fone. According to \engauge, we describe the
displacement from $\bR$ to $\bR+\dd\bR$ by an increment $\delta h(u)$ in
$h(u)$. But this displacement is only partly physical, i.e., only partly
normal to $\bR$, since it is along $\bn_0(u)$, which is not equal to the
normal $\bn(u)$ at $\bR(u)$. The actual normal displacement from $\bR$
to $\bR+\dd\bR$ is simply $\bn(u)\cdot\bn_0(u)\dl h(u)$, and we need
to replace $\dh\naive$ in \emnaive\ by $\CJ[h]\dh\naive$ where
\eqn\efpa{
\CJ[h]\equiv
\prod_{ij}\bigl(\bn(u_{ij})\cdot\bn_0(u_{ij})\bigr)\quad.}
The Jacobian $\CJ[h]$ is the first correction factor (the ``\fp\
determinant'') mentioned in the Introduction.

\ifx\bigans\answ
\ifigure\fone{Two surfaces $\bR,\bR+\dd\bR$ described in the normal gauge
associated to $\bR_0$.}{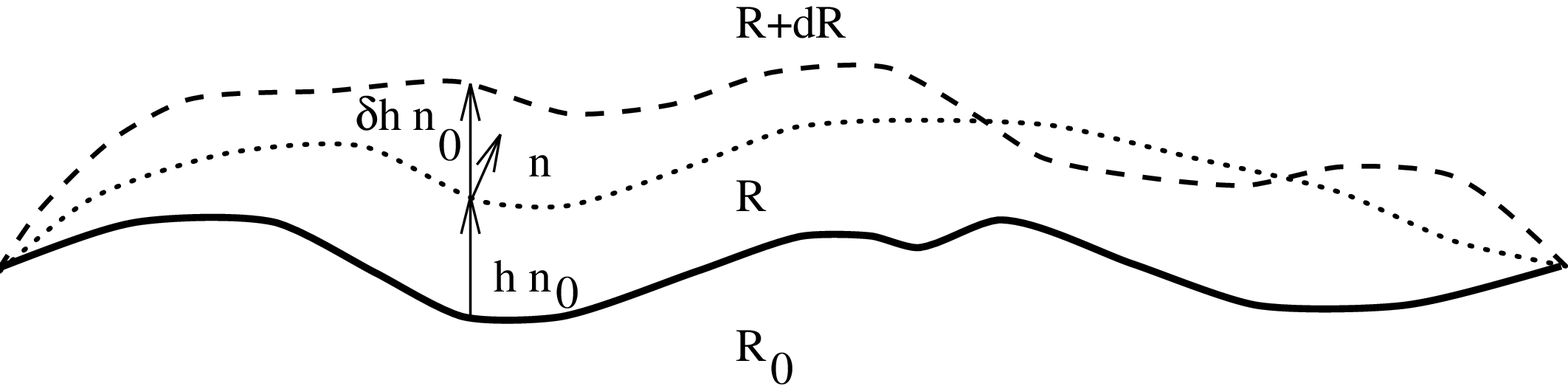}{1.5}
\else
\ifigure\fone{Two surfaces $\bR,\bR+\dd\bR$ described in the normal gauge
associated to $\bR_0$.}{surfaces.eps}{1.1}
\fi
Our derivation of \efpa\ was heuristic. Nevertheless the result agrees with
a more formal derivation which we will give in a moment, and the
heuristic discussion gives $\CJ[h]$ a very simple geometric meaning.
The formula \efpa\ also agrees with the approach based on metrics on
field space \bigchiral. For example, taking $\bR_0(u)$ to be {\it flat}
we have $\bn\cdot\bn_0=g^{-1/2}$, and $\CJ$ is the Monge gauge factor
found in \bigchiral. It is not unity, contrary to ref.~\DSMMS; while
it does not affect the one-loop renormalization of $\kappa_0$, it will
be crucial for $r$, as shown below and in \bigchiral. 

Our second derivation of \efpa\ is couched in language familiar from
gauge theories~\ref\Popov{V. Popov, {\sl Functional Integrals in
Quantum Field Theory and Statistical Mechanics} (Riedel, 1983).}%
. It is more precise than the motivation just given for
eqn.~\efpa, and again reproduces the answer given in \bigchiral.
We think that each derivation sheds light on (and increases our confidence
in) the final answer. 
In order to form a statistical
sum over distinct surfaces we imagine integrating over {\it all} parameterized
surfaces, with a correction factor to eliminate overcounting. Thus, we
write the partition function as
\eqn\epfa{
Z=\int[D\bR]\left(1\over\vol[\bR]\right)\,\ex{-\CH/T}\quad.}
Here $[D\bR]$
just describes the possible displacement of the individual mass-points on
the surface \fixup{and vol$[\bR]$ is a factor to be discussed below}%
. All we will need to know for the moment is that
the measure $[D\bR]$ must be coordinate-invariant, since no preferred
coordinate system is given on a fluid membrane; $[D\bR]$ must also be
invariant under spatial translations and rotations.

To finish specifying \epfa\ we need to specify the extra volume factor
vol$[\bR]$
appearing there. Starting from any surface $\bR(u)$ and applying all
possible reparameterizations we sweep out a subspace of all
parameterized surfaces.
The volume vol$[\bR]$ of this ``orbit'' may well depend on which surface we
start
with, but it cannot depend on how $\bR(u)$ is parameterized.

Since it proves inconvenient to include $\vol[\bR]$ explicitly in \epfa,
we now fix a coordinate choice by introducing two constraints on $\bR(u)$
at every point $u$:
$$
f_a(\bR(u))=0\qquad a=1,2\quad.
$$
For example, in normal gauge we require
\eqn\eqnleng{
f_a(\bR(u))\equiv
\lambda^{-2}\be_{a,0}(u)\cdot
(\bR(u)-\bR_0(u))=0\quad.}
We have introduced a factor
of $\lambda^{-2}$ \fixup{(see \elamb)}{} to make $f_a$ dimensionless.
Any length scale will
do here since our answers will be altogether independent of our choice
of $f_a$.
We now define the Faddeev-Popov determinant $\Delta_f[\bR]$ via
\eqn\efpb{1\equiv
\Delta_f[\bR]\cdot
\intprm [D\tilde\bR]\,\delta[f_a(\tilde\bR)]\quad.}
The integral is only over the space of $\tilde\bR$'s which are
reparameterizations of $\bR$; hence we also have
\eqn\evoli{
\vol[\bR]=\intprm [D\tilde\bR]\quad.}
To keep vol$[\bR]$, and hence \epfa, dimensionless we include two
factors of $\lambda\inv$ at every point in \evoli.
Since every $\tilde\bR(u)=\bR(\Omega(u))$ is related to $\bR$ by a
reparameterization, $u^a\mapsto\Omega^a(u)$, we can also write $\intprm
[D\tilde\bR]$ as $\int[D\Omega]\CD[\Omega,\bR]$, where formally
$\CD[\Omega,\bR]=\det\,{\dd(\bR(\Omega))\over \dd\Omega}$ is a
\fixup{Jacobian factor related to the}{} change of
variables. Let us now insert \efpb\ into \efpa\ to get
$$
Z=\int[D\bR]\left(1\over\vol[\bR]\right)\,
\ex{-\CH[\bR]/T}\Delta_f[\bR]\int
[D\Omega]\CD[\Omega,\bR]\delta[f_a(\bR(\Omega))]\quad.
$$
Since a reparameterization $\Omega$ has meaning independent of the surface
$\bR$ we apply it to, we can do the $\Omega$ integral last. Letting
$\tilde\bR=\bR(\Omega)$, we use the fact that $[D\bR]$, $\CH$, $\vol[\bR]$
and $\Delta_f$ are reparameterization invariant to rewrite $Z$ as
$$
Z=\int[D\Omega]\int[D\tilde\bR]
\left(1\over\vol[\tilde\bR]\right)\,
\ex{-\CH[\tilde\bR]/T}\Delta_f[\tilde\bR]
\delta[f_a(\tilde\bR)]\CD[\Omega,\bR]\quad.
$$
Also, the fact that $[D\bR]$ is coordinate-invariant means that small
displacements of $\Omega$ sweep out equal volumes no matter where on a
given orbit they act, \ie\ $\CD[\Omega,\bR]=\CD[\Omega,\tilde \bR]$.
Renaming $\tilde\bR$ as $\bR$ and using \evoli\ we thus get
\eqn\epfb{
Z=\int[D\bR]
\ex{-\CH[\bR]/T}\Delta_f[\bR]\delta[f_a(\bR)]\quad.}
The only configurations entering \epfb\ are those obeying
the gauge condition, as desired, but now we must compute $\Delta_f[\bR]$
from the definition \efpb.

While \epfb\ is quite general, we will now specialize to the normal gauge
\engauge\ associated to a surface $R_0$. The functional delta-function in
\epfb\
tells us we need only to examine $\Delta_f[\bR]$ for $\bR$ obeying the
normal-gauge condition; the delta-function in \efpb\ then tells us we need
only study $\tilde \bR$ infinitesimally close to $\bR$. Thus we write
$\tilde\bR$
in the form \ecoord, $\tilde\bR(u)=\bR(u)-\ep^a(u)\be_a(u)$. A
reasonable choice\foot{The nonlinear measure
correction to be discussed in the next section will not affect our
answer; see the Appendix.}
 for $[D\tilde{\bR}]$ near a given
$\bR$ is then
\eqn\erloc{
[D\tilde{\bR}]\equiv\prod_{ij}(\dd^3\tilde{\bR}(u_{ij})/\lambda^3)\quad.}
In \erloc\ we imagine that whatever coordinate system
we choose, a grid has been laid down with each point carrying constant
mass. 
Since each constituent molecule occupies a fixed physical area, we can
equivalently specify that each grid cell of size $(\Delta u^1,\Delta u^2)$
occupy equal areas in 3-space. (This, in turn, typically means that the
$\Delta u^a$ themselves are {\it not} all the same.) With this understanding
$[D\tilde{\bR}]$ is indeed coordinate-invariant.

Submitting $\bR$ to an infinitesimal coordinate transformation
$u^a\mapsto u^a+\ep^a(u)$,
we see that at each $u_{ij}$ $\bR(u_{ij})$ moves in
the tangent plane
; the measure
$\dd^3\tilde{\bR}(u_{ij})/\lambda^3$ in
\erloc, restricted to this plane, becomes
\eqn\emeae{
\lambda^{-2}|\be_1\times\be_2|\dd^2\epsilon(u_{ij})=
\lambda^{-2}\sqrt{g(u_{ij})} \dd^2\epsilon(u_{ij})\quad.}

We can now evaluate \efpb:
$$
\eqalign{
\Delta_f[\bR]^{-1}=\,&
\int\prod_{ij}(\lambda^{-2}\sqrt{g}\dd^2\epsilon(u_{ij}))
\delta^{(2)}(f_a(\bR(u_{ij}^a+\epsilon^a(u_{ij}))))\cr
=\,&\int\prod_{ij}(\lambda^{-2}\sqrt{g}\dd^2\epsilon(u_{ij}))
\det^{-1}[J_{ab}(u_{ij})]\delta^{(2)} (\epsilon^a(u_{ij}))\quad,\cr}
$$
where $J_{ab}(u)=\part f_a(\bR(u+\epsilon))/\part\epsilon^b=\lambda^2
\be_{a,0}\cdot\be_b$ in normal gauge using \eqnleng. Thus
\eqn\efpc{
\Delta_f[\bR]=\prod_{ij}{\det\be_a^0\cdot\be_b\over\sqrt{g}}\quad.}

To finish our job of getting an explicit expression for the partition
function, we now express $[D\bR]\delta[f_a(\bR)]$ appearing in \epfb\
in terms of the nonredundant height variable $h(u)$ appearing in
\engauge. We can conveniently express an arbitrary $\bR$ in terms of a
height change plus the rest:
\eqn\efng{
\bR(u)=\bR_0(u)+h(u)\bn_0(u)+\be_{a,0}(u)v^a(u)\quad.}
The calculation is not quite the same as the one leading to \emeae;
instead of infinitesimal reparameterizations $\epsilon^a$ from an
arbitrary normal-gauge surface $\bR$, now we have displacements
from the {\it fixed} reference surface $\bR_0$. Accordingly, instead
of \emeae\ we find
\eqn\emeav{
\dd^3\bR(u)=\sqrt{g_0(u)}\dd^2v\dd h\quad.}
Next, \engauge\ with \efng\ gives
$f_a(\bR)=\lambda^{-2}\be_{a,0}\cdot\be_{b,0}v^b$,
or with \efng\
$\delta^{(2)}(f_a(\bR))=\lambda^2\det^{-1}(\be_{a,0}\cdot\be_{b,0})
\cdot\delta^{(2)}(v^a)$. We can now do the $v^a$ integrals to get from \epfb\
to
\eqn\epfd{
Z=\int\Biggl[\prod_{ij}\lambda^{-1}dh(u_{ij})\Biggr]
\tilde\Delta \ex{-\CH/T}\quad,}
where
\eqn\efpd{
\tilde\Delta=\prod_{ij}{\det(\be_{a,0}\cdot\be_b)\over\sqrt{g}\sqrt{g_0}}=
\prod_{ij}\bn\cdot\bn_0(u_{ij})\quad.}
It is understood that $\CH$ and $\tilde\Delta$ are evaluated on the surface
given in terms of $h(u)$ by \engauge. The last step follows because
$\det(\be_{a,0}\cdot\be_b)=(\be_{1,0}\times \be_{2,0})\cdot(\be_1\times \be_2)=
\sqrt{g_0}\bn_0\cdot\sqrt{g}\bn$. Note that each factor in \epfd\ is
explicitly coordinate-invariant, in contrast with the analogous
formula in \DSMMS.

Eqns.~\epfd--\efpd\ are our final form for the partition function. We see
that \efpd\ indeed does agree with our heuristic formula \efpa. 
Specializing to Monge gauge, ${\bf n}_0=\hat z$, we recover the result
of \bigchiral, $\tilde\Delta=\prod_{ij}\iab g^{-1/2}$.

An alternate form for ${\tilde \Delta}$ can be obtained using
\eqn\einter{
\ev_b = \partial_b  \Rv = \ev_{b,0} + \partial_b ( h \nv_0) =
\ev_{b,0} - h \ev^c_0 K_{bc,0}+\pa_bh\nv_0 \quad,}
where $K_{bc,0}$ is the curvature tensor of the background surface.
{}From this we find
\eqn\efpe{
{\tilde \Delta} = \prod_{ij} {\det( g_{ab,0} - h K_{ab,0}) \over
\sqrt{g_0}\sqrt{g}} =\exp\sum_{ij}\bigl({-\half\partial^a h \partial_a h }
+\CO(h^4)\bigr)
\quad,}
which again differs slightly from the formula in \DSMMS. We see that
getting the correct $\sqrt g$ factors in the denominator is crucial to
avoid a spurious term linear in $h$.

Returning to the unhappy ending of sect.~3, we now see that the
Monge-gauge Faddeev-Popov factor
\eqn\edummy{\tilde\Delta=\exp\left[
-{1\over T}\sum_{ij}{T\over2}\log\bigl(1+\dhs
\bigr)\right]
=\exp\left[-{1\over T}\sum_{ij}{T\over2\iab}\dhs\right]}
 effectively
contributes a new term to \eHexpa, one which is already of order $T$.
Following the steps leading to \ernaive, we get the corrected version
\eqn\erFPtau{r\FP=\muo-{T\over2}\int^\Lambda\dq\left(1-{\muo q^2\over
\ko q^4+\muo q^2 }\right)\quad.}
On the other hand $\tilde\Delta$ cannot contribute to our calculation
of $\tau$ since being already of order $T$ we simply evaluate it at
$h=0$, where it equals one. We still have the problem that $r\FP\not=
\tau$.

\newsec{Nonlinear Measure}
\fixup{To see where we have erred, let us first give another
derivation that $r=\tau$. Note that there are many different Monge
gauges corresponding to projections to various reference
planes.\foot{\fixup{Recall that a Monge gauge is a normal gauge where the
reference surface $\bR_0$ in \fone\ is flat. Alternatively we could
imagine two different physical surfaces related by a rotation, both
viewed in the {\it same} Monge gauge.}}\
So far we
have used a reference plane coinciding with the plane of the }
frame holding the membrane, so
the equilibrium configuration is $\hb=0$, but we could instead choose
$\bR'_0$ tilted by a small angle $\psi$. We will always work to lowest
nontrivial order in $\psi$. In the new coordinate system the {\it
same} equilibrium surface is now described by $\hb'(u)=\psi
u^1$ (\tfig\ftwo);
furthermore the projected area has been foreshortened in one
direction, giving  $A_B'=\ab\cos\psi=\ab(1-\half\psi^2)$. Since the
two calculations
compute the same quantity we must have that
\eqn\egoodWI{\eqalign{\Ga[\ab;\hb=0]&\equiv\tau\ab\cr &=
\Ga[\ab-\half\psi^2\ab;\hb'(u)=\psi 
u^1]\cr& =\tau\ab-
\tau\half\psi^2\ab+ r\half\psi^2\ab\
.\cr}}
In the last step we wrote the $\CO(\bar h^{\prime2})$ term of $\Ga$ in
terms of $r
(\pa\hb')^2 =r\iab\psi^2$. Thus we again conclude $r=\tau$, contrary to
our still incomplete calculation.

\ifx\bigans\answ
\ifigure\ftwo{Tilting the reference surface foreshortens both the base
area and the coordinate cutoff.}{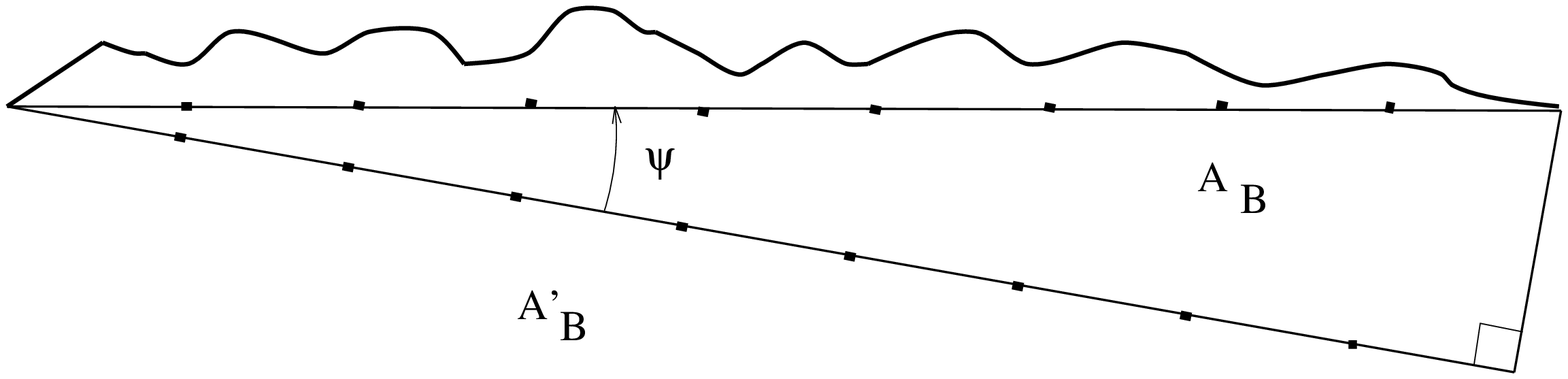}{1.4}
\else
\ifigure\ftwo{Tilting the reference surface foreshortens both the base
area and the coordinate cutoff.}{tilt.eps}{1}
\fi

\fixup{What have we done wrong?
Throughout our discussion of
the previous section we assumed that the measure $[D\bR]$, and its
reduction $[Dh]$, were invariant under changes of
parameterization of the surface. But the choice made in
\emnaive\ does not have this property. }
We chose the grid
points $\{u\ij\} $ evenly spaced in the coordinates $u^a$, which has
different meaning for different coordinate systems. In particular for
our second description we have a total of $A_B'/a^2$ degrees of
freedom, which is different from the true number $\ab/a^2$ of mass
points! More formally, eqn.~\erloc\ tells us to use grid cells of
constant {\it metric area} $\Delta u^1\Delta u^2=a^2/\sqrt g$, while
\emnaive\ specifies constant {\it coordinate area} $\Delta u^1\Delta
u^2=a^2$. 
\fixup{We can easily show that this error accounts for the remaining
discrepancy between $\tau$ and $r\FP$, eqns.~\etaur\ and \erFPtau.

Let us
denote by $\Ga\FP[\ab,\La^1,\La^2;\hb]$ the effective action calculated
with the correct \fp\ factor $\CJ$ but still using the
naive (coordinate-dependent) measure \emnaive. We momentarily allow an
anisotropic cutoff
$(\La^1,\La^2)$, which we write explicitly; at the end of the
discussion we will set $\La^1=\La^2=2\pi/a$. The foregoing discussion
implies that we may use $\Ga\FP$ if we remember to foreshorten the
{\it spacing} of constituents (increase $\La^1$)
as well as the base plane (\ftwo), so \egoodWI\ becomes
\eqn\ebadWI{\eqalign{\Ga\FP\bigl[\ab,{2\pi\over a},{2\pi\over a},\hb=0\bigr]
&=\tau\ab\cr&=
\Ga\FP\Bigl[\ab-\half\psi^2\ab,(1+\half\psi^2){2\pi\over a},{2\pi\over a};
\hb'(u)=\psi
u^1\Bigr]\cr}}
or
\eqn\emock{\tau\ab=\tau\ab-\half\tau\psi^2\ab+\half r\FP\psi^2\ab+
\half \psi^2{2\pi\over a}\left.\pd{\Ga\FP[\ab,\La^1,\La^2,\hb=0]}{\La^1}
\right|_{\La^1=\La^2=2\pi/a}\quad,}
since $r\FP$ is the \roughness  coefficient we calculated with the naive
measure. Thus we find 
\eqn\eFPtau{\tau=r\FP+\pd{\tau}{\log \La^1} \quad,}
which really is obeyed by our calculated expressions \etaur, \erFPtau.
}

We have thus identified the problem: the naive measure spreads mass
points evenly in coordinate space, not in real space. For flat
equilibrium surfaces ($h$ linear in $u^a$), the right prescription was
easy to find: we keep the grid points uniform but change their density
to get the RHS of \ebadWI. Clearly for nonflat $h(u)$ we will have to
be more clever, since here the cutoff is nonuniform and indeed
field-dependent: the desired measure $[Dh]$ is {\it nonlinear} in $h$.
Incidentally, the same problem we face here arises
again in the dynamics context \tclwc, where a naive calculation
gives the two-point function controlled by our $r\FP$.

We hinted at the answer in the Introduction: a change in the cutoff,
even a spatially-varying change, amounts to a local change in the
number of degrees of freedom. We can imagine passing from the desired
nonlinear
$[Dh]$ to the convenient $[Dh]\naive$ by a sort of decimation
procedure, where we integrate $\wgt$ over the discrepant degrees of
freedom to obtain $\ex{-\CH_\eff/T}$. The difference $\dl\CH=\CH_\eff -\CH$
is simply a counterterm; while in general it is a nonlocal functional
of $h$, still when we ask long-scale questions it may be replaced by a
renormalizable truncation, just as we did for $\CH$ itself.

We therefore seek a correction factor $\dl\CH$ with the property that
\eqn\enotJac{\hbox{``}\ [Dh]=\ex{-\dl\CH[h]/T}\dh\naive\ \hbox{''}\quad.}
This equation is in quotes because as it stands it makes no sense: we
cannot equate two measures with different numbers of degrees of
freedom. In particular $\dl\CH$ is not a Jacobian factor, as often
asserted\foot{We thank J. Distler for emphasizing this point%
.}, but rather is a {\it counterterm}. As such
it can and will depend on $\CH$ itself. Eqn.~\enotJac\ is not a change
of variables but rather an assertion about renormalization, namely
that each side when weighted by $\wgt$ yields identical moments of
$h$, at least for long wavelengths. The usual miracle of
renormalization is that adjusting just a few coefficients in $\dl\CH$
suffices to make all the moments agree.

What is $\dl\CH$? We have argued that it is renormalizable. Since it arises
from fluctuations it will be explicitly of order $T$, like the \fp\
correction, and so we will only need its form to $\CO(h^2)$ in order
to calculate $\la hh\ra$ to one loop. \fixup{(In the Appendix we show
how to do better than this.) } Dimension counting
says it can have at most four derivatives; translation invariance
requires at least two.
Furthermore $\dl\CH$ should
depend on $h$ only via the induced metric $g=\iab1+\dhs$, since it
reflects a change of cutoff controlled by this metric, and in
particular $\dl\CH$ should vanish at $h=0$, since here
$g_{ab}=\iab\dl_{ab}$ is the metric assumed in $\dh\naive$. Putting all
these arguments together we can only have
\eqn\edelH{\dl\CH=\half T\mul\int\du\dhs+\CO(h^4)\quad,}
where $\mul$ is some cutoff-dependent constant we are to find. (The reader
who is uncomfortable with the logic of this paragraph will find in the
Appendix  a more
deductive derivation of $\dl\CH$ obtained by matching two calculations
of $\Gah$.)

We remark that our $\dlh$ is analogous to the
``Liouville correction
factor'' familiar from the string-theory literature \DDK\PoStro. In
these papers the authors argue in analogy with the conformal anomaly
\Polbook\ that a covariantly cut-off measure (typically the measure
for the scale factor $\ex\sigma$ of an internal metric) ``equals'' a
measure cut off by a fixed metric $(g_0)_{ab}$ times a correction:
$$\hbox{``}\ [D\sigma]=\ex{\dlh[\sigma]/T}[D\sigma]_{g_0}\ \hbox{''}\quad,$$
where
\eqn\eLio{\dlh[\sigma]=T\int\sqrt{g_0}\du\bigl[\al(( \pa\sigma)^2+
R_0\sigma)+\mul\ex\sigma\bigr]}
and $R_0$ is the gaussian curvature of $g_0$. We are not using an
internal metric, but our \edelH\ is reminiscent of the last term of
\eLio\  (plus a constant).

The pleasant feature of our approach is that 
fixing the coefficient $\mul$  by comparison to a simple
situation 
then lets us compute $\Gah$ in any situation
. The simple situation we have in mind is the tilted but flat
reference surface appearing on the RHS of \ebadWI; $\dl\CH[\hb=\psi u^1]$
is supposed to reproduce the effect of cutoff-stretching, which we
introduced by hand via the last term of \emock. Thus we choose
$\mu_1=\pd{\tau}{\log\La^1}$, or
\eqn\emuone{\mul=
{T\over2}\int^\Lambda \dq\left[\log\Bigl(\muo q^2+\ko
q^4\Bigr){(a\lambda)^2\over2\pi T} + 1 - {\muo q^2\over\ko q^4+\muo q^2}
\right]\quad.}

Eqn.~\emuone\ fixes the effective Hamiltonian to order $h^2$. Since
\edelH\ is effectively just an additional contribution to the $q^2$
coefficient of $\la hh\ra$, we see that indeed the choice \emuone\
required by rotational invariance is also the choice which secures
$r=\tau$.  To
calculate four-point correlators, or two-loop corrections to the
effective couplings $\mu_\eff,\kappa_\eff$ we would need also the
$\CO(h^4)$ terms of \edelH; \fixup{see the Appendix}.

\optional{We can also easily see how to generalize our analysis to include
fields representing in-plane order on the surface: for example, if
$\theta(u)$ is a tilt-angle field then we need a counterterm to relate
$[D\theta]$ to $[D\theta]\naive$. Since $\theta$ does not itself
affect the induced metric, this calculation is just the usual
conformal anomaly \Polbook, which makes further contributions to $\al$
and $\mul$.}

\newsec{Conclusion}

If we have explained ourselves clearly the reader may find our
analysis somewhat tautological. After all, we uncovered a paradox,
only to eliminate it with a   mysterious counterterm chosen solely for
that purpose! Let us now comment on the real content of our analysis.

We have studied the construction of statistical ensembles of random
surfaces made out of fixed-size constituents. Such ensembles are good
models for the long-scale behavior of fluid bilayer membranes, though
it is straightforward to introduce in-plane order as
well. The correct statistical weight for such an ensemble involves a
subtle nonlinear measure $\dh$ not directly suited to diagrammatic
perturbation theory. We have argued that this measure may be replaced
by a simple one if we incorporate two correction factors
into an effective Hamiltonian.

We have given simple geometrical interpretations to the two correction
factors, as follows: the \fp\ term describes how a change $\dl h$ of
the normal displacement may not itself be a normal displacement
(\fone). The Liouville term describes how the molecules of a surface,
projected down to a reference plane, seem to crowd together in regions
where the surface is tilted relative to that plane (\ftwo). These
simple pictures, plus some general arguments, enabled us to write down
the complete form of the effective Hamiltonian up to terms of order $T$.
This accuracy sufficed to show these corrections are
crucial to secure the covariant form of the effective action argued on
general grounds in sect.~2. Besides being satisfying, the geometrical arguments
made it easier to get the right answers than other more formal arguments.

Our argument feels tautological because we used one consistency
condition ($r=\tau$) to fix one counterterm ($\mul$). In the appendix
we sketch how to fix the rest of $\dlh$ to one loop accuracy, but to go
farther we would need to fix its form in advance by imposing some sort
of Ward identity, perhaps along the lines of \PoStro.
 This task we leave for future work.

\vskip1truein \leftline{\bf Acknowledgements}
{\frenchspacing\noindent
We would like to thank J. Polchinski, A.M. Polyakov, and especially
J. Distler for conceptual help.} P.N. thanks the Aspen Center for
Physics for hospitality while some of this work was done.
This work was supported in part by NSF
grants PHY88-57200, DMR-91-20668 and DMR-91-22645, and the Donors of the
Petroleum Research Fund.

\appendix{A}{Some Calculations}

In Section 4 we used rotational invariance to fix the value of part of
the cutoff-stretching factor; here we present another approach.  Our
strategy is to compute the effective action in two different ways;
demanding that the results of these calculations agree determines the
cutoff-stretching factor to one-loop accuracy, i.e. leading nontrivial
order in $T$.

We follow the standard procedure for computing the one-loop effective
action, $\Gah$ \Polbook.  We write $h$ as $\hb + \delta h$, expand
the energy functional to quadratic order in $\delta h$, and drop the
term linear in $\delta h$.  Then we do the gaussian integral over
$\delta h$ to get the one-loop contribution $\delta\Gah$ to the
effective action.  Our key observation is that for gaussian integrals
we don't really need a measure on field space, since in the end we
just compute functional determinants, and for this we need only a {\it
metric} on fluctuations $\delta h$ about the given $\hb$.
Equivalently, to one-loop accuracy we make no error if we replace the
full measure $[D\delta h]$, which is nonlinear in $\delta h$, by the
cut off measure
\eqn\emspacing{[D\delta h]_{\bar g} = \prod_{ij,\gb} {\dd\delta h(u\ij
)\over\lambda}\quad,}
\fixup{where the notation ``$ij,\gb$'' means that the }
grid points are spaced uniformly in terms of the distance given by the
background metric $\bar g_{ab} = \delta_{ab} + \pa_a\hb \pa_b\hb$,
with density  $a^{-2}=(\La/2\pi)^2$.
\fixup{As we emphasize below this does {\it not} necessarily mean the
points are spaced uniformly in any given {\it coordinate} system. }
The measure \emspacing\ is independent of $\delta h$, unlike the
desired $[D\delta h]$, but both correspond to the same metric on
fluctuations about $\hb$.

Stated succinctly, in our first calculation we cut off using $\bar g$
and {\it no} cutoff-stretching correction (though we do need the \fp\
factor as usual).


It is a bit unsatisfying to have to use a different measure
\emspacing\ for every background $\hb$.  Moreover, while we succeeded
in determining the free energy for various backgrounds $\hb$, this is
not quite the same as finding the {\it correlations} of height; for
the latter, the $\dl h$-dependence of the true measure omitted from
\emspacing\ will matter.  So in our second calculation we replace
$[D\dl h]$ not by $[D\dl h]_{\bar g}$ but by
\eqn\emmm{\ex{-\dlh[h]/T}[D\dl h]_{g_0}.}
\fixup{Here $[D\dl h]_{g_0}$ is defined analogously to \emspacing\
except that now $g_{0,ab}=\dl_{ab}$ is the usual {\it flat}
metric}, so there is no explicit dependence on the background $\hb$:
we have one measure from which we compute the various $\Gah$.
\fixup{Since $\go$ is flat, in this calculation the grid points really
are spaced uniformly in the coordinates $u^a$. } On the
other hand, now we have to expect a correction $\dlh$ as discussed in
the text.  We will find $\dlh$ to one loop by recomputing $\Gah$ and
comparing to the first calculation; it turns out that a single
universal $\dlh$ secures agreement with all
the different $\Gamma$'s in the first calculation, which used a
different measure for each $\hb$.  With $\dlh$ in hand we can then
proceed to the calculation of $\la hh\ra$.

We begin our calculations by quoting some Monge gauge formulas. 
We have the following expressions for the volume element, the inverse
metric, and the mean curvature \Kla:
\eqn\eve{\sqrt{g}\du = \sqrt{1 + \dhs}\du\quad,}
\eqn\einvmet{g^{ab} = \dl_{ab}-{{\pa_a h\pa_b h}\over{1+\dhs}}\quad,}
\eqn\emc{{\rm Tr}K = {-\pa^2 h\over\sqrt{1+\dhs}} + {\pa_a\pa_b h \pa_a
h \pa_b h\over (1+\dhs)^{3/2}}\quad.}
The energy functional \eCaHeb\ becomes
\eqn\eHex{\eqalign{\CH={\ko\over{2}}&\int\du
\left\{
{(\partial^2h)^2\over \sqrt{1+\dhs}}
-{2\partial^2h\pa_a h\pa_b h\pa_a\pa_b h\over (1+\dhs)^{3/2}}
+{(\pa_ah\pa_bh\pa_a\pa_bh)^2\over(1+\dhs)^{5/2}}
\right\} \cr
&+\muo\int\du\sqrt{1+\dhs}\quad.\cr
}}
We work to all orders in the height.

\def\lfr#1#2{{#1\over#2}} 
We are now ready for the first of our two calculations. Following the
above program, we replace $h$ by $\bar h+\dl h$ and retain the terms
quadratic in $\dl h$. In the following we drop all terms that do not
contribute to the frame tension renormalization, \ie\ \fixup{terms
with more
derivatives than $\hb$'s, though it's not much harder to retain them. }
We then have the quadratic bit of $\CH$,
\eqn\ehtwoa{\eqalign{\CH^{II}={\ko\over{2}}&\int\du \rgb \dl h\left\{
\lfr1{\bar g}\partial^4
-{2\pa_a\bar h\pa_b\bar h\over\bar g^2}\partial^2\pa_a\pa_b
+{\pa_a\hb\pa_b\hb\pa_c\hb\pa_d\hb\over\bar g^3}\pa_a\pa_b\pa_c\pa_d
\right\}\dl h\cr
&+{\muo\over 2}\int\du\rgb\dl h\left\{
-{1\over\gb}\pa^2+{\pa_a\hb\pa_b\hb\over\gb^2}\pa_a\pa_b\right\} \dl h\quad.\cr
}}
Note that we have pulled out a factor of $\rgb$ so that upon
discretizing this integral \ehtwoa \ the grid points are spaced with
uniform density according to the metric $\gb_{ab}$ (see \emspacing).
The result of this gaussian integration is
\eqn\eGa{\dl \Gamma_{\rm gaussian}[\hb]=
\lfr T2\Tr_\gb\log\Bigl({\ko\lambda^2a^2\over2T\pi}\Bigr){1\over\gb}\left[
(\gb^{ab}\pa_a\pa_b)^2+{\muo\over\ko}(\gb^{ab}\pa_a\pa_b)\right]\quad.}
\fixup{In this equation the symbol ``$\Tr_\gb$'' means that we
evaluate the trace of the following operator using a cutoff equivalent
to the point spacing specified in \emspacing. In momentum space this
means that we find a basis of eigenmodes of the covariant laplacian
$\Db={1\over\rgb}\pa_a{\bar g}^{ab}\rgb\pa_b$, then sum over all
modes with eigenvalues less than $(\La)^2$. This prescription is
coordinate-invariant; if $\gb_{ab}=\dl_{ab}$ then it just amounts to
integrating over a circular Brillouin zone as usual. }
We have
simplified \eGa\ by factoring out the constants and one power of
$\gb\inv$.

Eqn.~\eGa\ is easy to work with because
we can now write $\gb^{ab}\pa_a\pa_b$ as $\Db$ plus terms with
more derivatives than $\hb$'s, which we are dropping:
\eqn\eGb{\dl\Gamma_{\rm gaussian}[\hb]=\lfr T2\Tr_\gb\log{\ko(\lambda
a)^2\Lambda^4\over 2\pi T}-\lfr T2\Tr_\gb\log\gb
+\lfr T2\trg\left(\Db^2-{\muo\over\ko}\Db\right)\La^{-4}
\quad.}
To this expression we must add the contribution from the Faddeev-Popov
factor $\prod_{ij,\gb}\gb^{-1/2}$, which we write as $\exp\Bigl[-\lfr1T
\bigl(\half T\Tr_\gb\log\gb\bigr)\Bigr]$:
\eqn\eEAa{\dl\Ga[\hb]=\lfr T2\Tr_\gb\log{\ko\lambda^2a^2\Lambda^4\over2\pi T}+
\lfr T2\Tr_\gb\log\gb-\lfr T2\Tr_\gb\log\gb+
\lfr T2\Tr_\gb\log\left({\Db^2\over\Lambda^4}-{\muo\over\ko\Lambda^4}\Db
\right)\quad.}
Note that the noncovariant \fp\ factor exactly cancels with the
noncovariant factor of $\gb\inv$ from the expansion of the energy
functional, leaving us with a contribution to the effective action
which is covariant to all orders in the height $\hb$. It is easy to
evaluate the first trace:
\eqn\etracea{\trg{\ko\lambda^2a^2\Lambda^4\over 2\pi T}=\log
{\ko\lambda^2a^2\Lambda^4\over 2\pi T}\sum_q1\quad.}
\fixup{The sum is just the total number of mass points $a^{-2}\int\du\rgb$,
which we will write as}
\eqn\edummy{=\int\du\rgb\int^\Lambda\dq\log
{\ko\lambda^2a^2\Lambda^4\over 2\pi T}\quad.}
\fixup{To make the last term of \eEAa\ tractable we will now
specialize to a special class of backgrounds $\hb$ for which we know
the \ei modes of $\Db$ explicitly, namely $\hb(u)$ linear in $u^a$.
Examining this case suffices to let us extract the coefficients of all
the $(\pa\hb)^n$ terms.\foot{{\fixup Alternately we could simplify the sum by
considering only the contribution from a thin shell in momentum space. }}\
For such a $\gb_{ab}$ we have
\eqn\etraceb{\eqalign{\trg\bigl(&\Db^2-{\muo\over\ko}\Db\bigr)\Lambda^{-4}
\cr&=\int\du
\int_{q_aq_b\gb^{ab}<\Lambda}\dq\log\left(
(q_aq_b\gb^{ab})^2
+{\muo \over\ko}q_aq_b\gb^{ab}\right)\La^{-4}\quad.\cr
}}
After a linear transformation of integration variables we get
$$=\int\du\rgb\int_{k^2<\Lambda}\dk\log\left(k^4+
{\muo\over\ko}k^2\right)\La^{-4}\quad.$$
}
The final result for the renormalized frame tension is thus
\eqn\etension{\tau=\muo+\lfr T2\int^\Lambda\dq\log
{\lambda^2a^2\over 2\pi T}(\ko q^4+\muo q^2)\quad,
}
which is indeed the contribution to \etaur\ due to the modes in
question, and agrees with \DaLe.\foot{We have corrected an error in
the formula following (B.2) in \bigchiral; it had an extra factor of
$\rgb$ (cf. \etraceb).}

We now turn to our second derivation described above: we compute the
part of the effective action involving the renormalized frame tension
 using
the measure cut off by a fixed metric $\goab=\dl_{ab}$. To space the
grid points
with uniform density according to $\goab$, we write \ehtwoa\ as
\eqn\ehtwob{\CH^{II}=\lfr\ko 2\int\du\rgo\dl h{1\over\sqrt{\go\gb
\mathstrut}}\left(\Db^2-{\muo\over\ko}\Db\right)\dl h\quad.
}
\fixup{While the laplacian $\Db$ associated to $\gb$ (not $\go$)
again appears in
\ehtwob, now our gaussian integral is cut off using a basis of \ei
modes of the {\it flat} $\pa^2$:}
\eqn\eGbb{\dl\tilde\Ga_{\rm gaussian}=\lfr T2\trgo
{\ko\lambda^2a^2\over 2\pi T}{1\over\sqrt{\go\gb\mathstrut}}
\Bigl(\Db^2-{\muo\over\ko}\Db
\Bigr)\quad.
}
To \eGb\ we must add the \fp\ factor and the cutoff-stretching factor.
Both of these are already $\CO(T)$, so to one loop we simply evaluate
them at $\hb$:
\eqn\eEAb{\dl\Ga=\lfr T2\trgo {\ko\lambda^2a^2\over 2\pi T}
{1\over\sqrt{\go\gb\mathstrut}}
\Bigl(\Db^2-{\muo\over\ko}\Db\Bigr)+\dl\CH[\hb]+\lfr T2\trg\gb\quad.
}

It is now easy to see what factor $\ex{-\dl\CH[h]/T}$ belongs in \emmm.
$\dl\CH[h]$ should be the functional of $h$ which, when evaluated at
$\hb$, reconciles \eEAb\ with \eEAa:
\eqn\eCSF{\dl\CH[h]=\lfr T2\bigl(\Tr_g-\Tr_\go)\log
{\ko\lambda^2a^2\over 2\pi Tg}\Bigl(\Delta^2-{\muo\over\ko}\Delta\Bigr)
-\lfr T2\trgo\left({g\over\go}\right)^{1/2}\quad.
}
We have written the correction factor this way to emphasize how at
least part of it (the first term of \eCSF) is clearly a result of
decimation.
\fixup{%
As in \etraceb\ we can write the first term of \eCSF\ as the
difference of the integral over an elliptical Brillouin zone minus the
integral over a smaller, circular, zone with the same density of
points $\ab$ (\tfig\fthree). Thus
the first term of \eCSF\ may be written as
\eqn\ebz{\left(\int\du\int_{BZ'}\dq -
\int\du\int_{BZ}\dq\right)\log{\lambda^2a^2\over 2\pi
T}\left(\ko(g^{ab}q_a q_b)^2 + \muo g^{ab}q_a q_b\right)}
where $q$ is in $BZ'$ if $q_a q_b g^{ab} < \Lambda^2$ and $q$ is in
$BZ$ if $q^2 < \Lambda^2$.} 

\ifx\bigans\answ
\ifigure\fthree{The origin of the cutoff-stretching counterterm as a
difference between two Brillouin zones. The stretched cutoff $\La'$
equals $\La/\cos\psi$ if the background surface $\hb$ is a flat plane
tilted by an angle $\psi$ relative to the reference plane.}{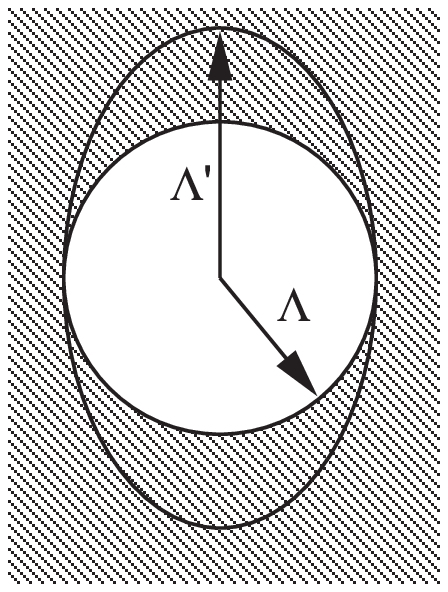}{1.5}
\else
\ifigure\fthree{The origin of the cutoff-stretching counterterm as a
difference between two Brillouin zones. The stretched cutoff $\La'$
equals $\La/\cos\psi$ if the background surface $\hb$ is a flat plane
tilted by an angle $\psi$ relative to the reference plane.}{brillo.eps}{1.1}
\fi

We can recast \eCSF\ as follows:
\eqn\edummy{\eqalign{\dl\CH[h]=&{T\over2}\int\du\rgb\int\dq\log
{\lambda^2a^2\over2\pi T}\bigl(\ko q^4+\muo q^2\bigr)\cr
&-{T\over2}\int\du\int\dq\log {\lambda^2a^2\over2\pi T}
\left(\ko(g^{ab}q_a q_b)^2 + \muo g^{ab}q_a q_b\right)\cr
&-{T\over2} \int\du\rgb\int\dq\log\gb\cr}
}
In this form we can readily see that the second line contributes a
term to the free energy cancelling the contribution from Feynman
diagrams, the third line cancels the contribution from the \fp\
determinant, and the remaining first line is covariant as required.
{}From this form one can show that all the terms of $\dl\CH$
proportional to $(\pa h)^n$ are just what are needed to satisfy the
Ward-like identity \egoodWI\ to all orders in the angle $\psi$.

Finally, we can expand
\eCSF\ in terms of $h$; the first couple of terms are
\eqn\eexpand{\eqalign{\dl\CH[h]=\lfr T2\int\du\int^\Lambda\dq
&\left[\log{\lambda^2a^2(\ko q^4 + \muo q^2)\over 2\pi T}+1
-{\muo q^2\over\ko q^4+\muo q^2}\right]\cdot\half\dhs\cr&+\CO(h^4)\quad,\cr
}}
in agreement with \emuone.

To conclude our analysis, we finally compute the last correction to
the \roughness  coefficient $r$. As mentioned earlier, the $\CO(h^2)$
term of \eexpand\ leads to a new correction to $\la hh\ra$, not
accounted for in the steps leading to \erFPtau. Since this term is
$\CO(T)$, it gives a new tree Feynman graph, which when added to the 1-loop
part of \erFPtau\ gives $r=\tau$. 

\footatend\bigskip\immediate\closeout\rfile\writestoppt
  \baselineskip=22pt\centerline{{\bf References}}\bigskip{\frenchspacing%
  \parindent=20pt\escapechar=` \input refs.tmp\vfill\eject}\nonfrenchspacing
 \vfill\eject\immediate\closeout\ffile{\parindent40pt
 \baselineskip22pt\centerline{{\bf Figure Captions}}\nobreak\medskip
 \escapechar=` \input figs.tmp \vfill\eject
}

\bye